\def\MPL #1 #2 #3 {Mod.~Phys.~Lett.~{\bf#1},\  #2 (#3)}
\def\NPB #1 #2 #3 {Nucl.~Phys.~{\bf#1},\  #2 (#3)}
\def\PLB #1 #2 #3 {Phys.~Lett.~{\bf#1},\  #2 (#3)}
\def\PR #1 #2 #3 {Phys.~Rep.~{\bf#1},\ #2 (#3)}
\def\PRD #1 #2 #3 {Phys.~Rev.~{\bf#1},\  #2 (#3)}
\def\PRL #1 #2 #3 {Phys.~Rev.~Lett.~{\bf#1},\  #2 (#3)}
\def\RMP #1 #2 #3 {Rev.~Mod.~Phys.~{\bf#1},\  #2 (#3)}
\def\ZP #1 #2 #3 {Z.~Phys.~{\bf#1},\  #2 (#3)}
\def\IJMP #1 #2 #3 {Int.~J.~Mod.~Phys.~{\bf#1},\  #2 (#3)}

\def\lsim{\mathrel{\raise.3ex\hbox{$<$\kern-.75em\lower1ex\hbox{$\sim$}}}}
\def\gsim{\mathrel{\raise.3ex\hbox{$>$\kern-.75em\lower1ex\hbox{$\sim$}}}}
\def\@versim#1#2{\vcenter{\offinterlineskip
        \ialign{$\m@th#1\hfil##\hfil$\crcr#2\crcr\sim\crcr } }}

\def\slash#1{#1\hskip-6pt/\hskip2pt}
\def\ptmiss{\slash p_T}

\documentstyle[preprint,aps]{revtex}
\begin{document}

\draft

\preprint{$
\begin{array}{l}
\mbox{UB--HET--96--03}\\[-3mm]
\mbox{UCD--96--32}\\[-3mm]
\mbox{MADPH--96--967}\\[-3mm]
\mbox{October~1996} \\   
\end{array}
$}

\title{Implications of $\mu\nu \gamma \gamma$ Production 
on Precision Measurements of the $W$ 
Mass\footnote{To appear in the 1996 Snowmass Proceedings:
New Directions for High Energy Physics.}
\thanks{Work 
supported in part by the Department of Energy, the National Science
Foundation, and by the Davis Institute for High Energy Physics.}}

\author{U. Baur$^a$\footnote{e-mail: baur@ubhep.physics.buffalo.edu},
 T. Han$^b$\footnote{e-mail: than@ucdhep.ucdavis.edu},
 R. Sobey$^b$\footnote{e-mail: rsobey@ucdhep.ucdavis.edu},
 and D.~Zeppenfeld$^c$\footnote{e-mail: dieter@pheno.physics.wisc.edu} }

\address{
$^a$Department of Physics,
State University of New York, Buffalo, NY  14260, USA\\
$^b$Department of Physics,
University of California, Davis, CA 95616, USA\\
$^c$Department of Physics,
University of  Wisconsin,  Madison, WI 53706, USA\\}

\maketitle

\begin{abstract} 
The process $p \bar p \rightarrow \mu^- \bar \nu \gamma \gamma$ is
calculated including finite lepton mass effects. Implications for 
precision measurements of the $W$ mass at the Tevatron are discussed.

\end{abstract}

\section{Introduction}
A precise measurement of the $W$ boson mass ($M_W$)
is of fundamental importance in testing the standard
model (SM) and in constraining new physics. Future experiments at
LEP2~\cite{lep} and the Tevatron~\cite{tev2000,ewk_group} aim at a 
precision of about 40~MeV and 20~MeV for $M_W$, respectively. 
In order to be able to achieve a precision of $\cal O$(10~MeV)
for $M_W$ in hadron collider experiments, it will be necessary to 
control not only the higher order QCD
corrections, but also the electroweak radiative corrections. 

At ${\cal O}(\alpha^n)$, $W$ decay with collinear emission of photons 
from a final state charged lepton of mass $m_\ell$ gives rise to terms 
which are proportional to $(\alpha/\pi)^n\log^n(M_W^2/m_\ell^2)$, in
$n$-photon exclusive rates.  These terms can have 
significant effects in determining $M_W$. At ${\cal O}(\alpha)$,
photon emission shifts the $W$ mass by~\cite{BKW}
\begin{equation}
\Delta M_W\approx -{\pi\beta\over 8}\,\Gamma_W,
\end{equation}
where $\Gamma_W$ is the total width of the $W$ boson, and
\begin{equation}
\beta={\alpha\over\pi}\left(\log{M_W^2\over m_\ell^2}-1\right).
\end{equation}
When finite detector resolution effects are included, $\Gamma_W$ is
replaced by the uncorrected experimental width. For CDF, the resulting
shift is $-168$~MeV ($-65$~MeV) in the muon (electron)
channel~\cite{cdfmw}. Similar results are found in the D\O\ $W$ mass
analysis~\cite{D0Wmass}. In the electron case, understanding the exact
amount of energy lost in photon bremsstrahlung is also important to
determine the $E/p$ distribution, which is used by CDF to determine the
energy scale of the central electromagnetic calorimeter~\cite{cdfmw}.
When aiming for a precision of ${\cal O}(10$~MeV) in $M_W$, it is 
therefore important to
control the effects of multiple photon radiation in $W$ events. 

Here, we present a calculation of two photon radiation in the 
$W\to\mu\nu$ channel at the tree level, including the finite mass of the
muon which regulates the collinear singularity originating from photons
radiated off the final state muon. Initial and final state
bremsstrahlung, together with finite $W$ width effects are taken into
account in our calculation. 

\section{Calculational Details}

The cross section for $p \bar p \rightarrow \ell^- \bar \nu \gamma \gamma$ 
has been calculated in the limit of massless fermions in 
Ref.~\cite{hsprd} and~\cite{bhksz}, using the helicity amplitude technique of 
Ref.~\cite{hz}. To evaluate the matrix elements for a massive final 
state charged lepton, we use the MADGRAPH~\cite{MadGraph} package, which
automatically generates the SM matrix elements in HELAS format~\cite{HELAS}.

However, in order to maintain electromagnetic gauge invariance in 
presence of finite $W$ width effects, the $W$ propagator and the
$WW\gamma$ and $WW\gamma\gamma$ vertex functions in the amplitudes
generated by MADGRAPH have to be modified~\cite{bhksz,bzprl}. Finite
width effects are included by resumming the imaginary part of the $W$
vacuum polarization, $\Pi_W(q^2)$. The transverse part of $\Pi_W(q^2)$
receives an imaginary contribution 
\begin{equation}
{\rm Im}\,\Pi^T_W(q^2)=q^2{\Gamma_W\over M_W}
\end{equation}
while the imaginary part of the longitudinal piece vanishes. 
The $W$ propagator is thus given by
\begin{eqnarray}
\label{EQ:PROP}
D^{\mu\nu}_W(q)= \frac{-i}{q^2 - M^2_W + iq^2 \gamma_W}
\left[ g^{\mu\nu} - \frac{q^\mu q^\nu}{M^2_W}
( 1 + i \gamma_W ) \right],
\end{eqnarray}
with
\begin{equation}
\gamma_W={\Gamma_W\over M_W}.
\end{equation}
A gauge invariant expression for the amplitude is then obtained by
attaching the final state photons to all charged particle propagators,
including those in the fermion loops which contribute to $\Pi_W(q^2)$.
As a result, the triple and quartic gauge vertices
($\Gamma^{\alpha\beta\mu}_0$ and $\Gamma^{\alpha\beta\mu\rho}_0$ ) are 
modified~\cite{bhksz,bzprl} to
\begin{eqnarray}
\label{EQ:VERTEX}
\Gamma^{\alpha\beta\mu} &=& \Gamma^{\alpha\beta\mu}_0 ( 1 + i \gamma_W),
\\
\Gamma^{\alpha\beta\mu\rho} &=& \Gamma^{\alpha\beta\mu\rho}_0 ( 1 + 
i \gamma_W ).
\end{eqnarray}

Due to the mass singular terms which arise from the $\mu\gamma$
collinear region, it is nontrivial to numerically compute the
$\mu\nu\gamma\gamma$ cross section. To optimize the Monte Carlo integration,
we choose the logarithms of the invariant masses of the $\mu\gamma_1$, 
the $\mu\gamma_2$, and
the $\mu\gamma_1\gamma_2$ system as integration variables:
\begin{eqnarray}
dPS_4 \sim d\hat s \ d\log m(\mu\gamma_1\gamma_2) \
d\log m(\mu\gamma_1) \ d\log m(\mu\gamma_2).
\end{eqnarray}

The resulting cross section is checked for gauge invariance, and is 
numerically stable when the photons are collinear with the muon. 

For $p\bar p\to e\nu\gamma\gamma$, due to the smaller
electron mass, the singularities for ${\it both}$ photons and the electron
being collinear, are much more severe than in the muon case. The method used
in $p\bar p\to\mu\nu\gamma\gamma$ to perform the phase space integration
is not sufficient to obtain a stable numerical answer for 
$p\bar p\to e\nu\gamma\gamma$. 

\section{Results and Discussion}

For our numerical simulations we use the MRSA set of parton
distributions~\cite{mrsa}, $M_W = 80.22$~GeV, and $\alpha=1/128$. 
The factorization scale is fixed to the parton center of
mass energy, $\sqrt{\hat s}$. To simulate detector response, we impose the
following acceptance cuts on transverse momenta and pseudorapidities:
\begin{equation}
\label{EQ:CUTS}
p_T(\mu) > 25\ {\rm GeV}, \quad  \ptmiss > 25\ {\rm GeV},
\end{equation}
\begin{equation} 
|\eta(\mu)| <1.0, \quad |\eta(\gamma)| <3.6.
\end{equation}
If the two photons have a separation smaller than a critical value 
$\Delta R(\gamma,\gamma)=R_c$, they cannot be discriminated in the
detector, and thus will
be treated as one photon. Taking the typical cell size of 
an electromagnetic calorimeter as a criterion~\cite{ecal}, we choose
$R_c=0.14$ in the following. The results of our simulations are listed
in Table~\ref{T:ONEGAMMA}. 

When no separation requirement is imposed on the muon and
the photon,  approximately 13\% (5.7\%) of all $W\to\mu\nu$ events 
contain one photon with a minimum photon transverse energy of 
$E_T^0(\gamma)=0.1$~GeV ($E_T^0(\gamma)=1$~GeV)~\cite{bzprl} at ${\cal
O}(\alpha)$. 
About 0.8\% (0.13\%) of all events
contain two isolated photons with transverse energies larger than 
$E_T^0(\gamma)=0.1$~GeV ($E_T^0(\gamma)=1$~GeV). If photons are required
to be isolated from the muon ($\Delta R(\mu,\gamma)>0.7$), about 3.8\% 
of the $W\to\mu\nu$ events contain one photon with $E_T(\gamma)>0.
1$~GeV, but only 0.07\% have two photons for the same $E_T(\gamma)$
threshold (see Table~\ref{T:ONEGAMMA}b). 
For larger photon $E_T$
thresholds, the fraction of two photon events drops very rapidly. 

Although the two-photon tree level calculation presented here is not a 
full next-to-next to leading order calculation of $W^\pm$ production,
we can still gain some important information on multiphoton radiation. 
Knowing the cross section for the lowest order process, $\sigma_0$, and
for ${\cal O}(\alpha)$ one photon emission, 
$\sigma_1$, one can estimate the probability, $P(n)$, for radiating 
$n$ photons.  Neglecting the dependence of the soft photon
emission rate on the (hard) quark and lepton configuration, the 
probability is given by~\cite{poisson}
\begin{eqnarray}
\label{EQ:PROB}
P(n) = \frac{<n>^n}{n!}e^{-<n>},
\end{eqnarray}
where $<n> = \sigma_1/\sigma_0$. Comparison with Table~\ref{T:ONEGAMMA}
shows that Eq.~(\ref{EQ:PROB}) provides a useful guideline for multiple
photon radiation in $W$ events if the minimum photon $E_T$ threshold is
smaller than about 3~GeV.

In summary, we have calculated two photon bremsstrahlung in $W\to \mu\nu$
events at hadron colliders. Our calculation takes both initial and final
state photon radiation, and finite $W$ width effects into account. We
also incorporate finite muon mass effects, which makes our calculation
valid over the entire phase space region. Approximately 0.8\% of all
$W\to\mu\nu$ events are found to contain two isolated photons with
transverse energy $E_T(\gamma)>0.1$~GeV, which roughly corresponds to
the tower threshold of the electromagnetic calorimeter of CDF and
D\O~\cite{ecal}. Multiple photon bremsstrahlung thus is expected to have
a non-negligible effect on the $W$ mass extracted from experiment. 
Unfortunately, our calculation at present does not produce stable
numerical results in the $W\to e\nu$ case, due to the more severe nature
of the collinear singularities in the electron case. 
It, therefore, should only be viewed as a very first step 
towards a more complete understanding of multiple photon radiation in 
$W$ boson events. 


\begin{table}[h]
\begin{center}
\caption[]{Fraction of $W \rightarrow \mu \nu$ events (in percent)
containing one or two photons at the Tevatron ($\sqrt{s}=1.8$~TeV),
(a) with no isolation cut between the photons and the muon imposed, (b)
requiring $\Delta R(\mu,\gamma)>0.7$. In the two photon case, the two
photons are required to fulfill the condition $\Delta R(\gamma,\gamma)
>0.14$. Fractions are obtained by normalization with respect to 
$\sigma_0 =  0.346$ nb.}
\label{T:ONEGAMMA}
\begin{tabular}{lcc}
$ $&(a) no isolation cut&$ $\\
\hline
\hline
$E^0_T(\gamma)$ & $W\to\mu\nu\gamma$ & 
$W\to\mu\nu\gamma\gamma$ \\
\hline
$0.1 $ & $13.4$  & $0.77$ \\
$0.3 $ & $9.60$  & $0.39$ \\
$1.0 $ & $5.67$  & $0.13$\\
$3.0 $ & $2.62$  & $0.024$\\
$10.0$ & $0.32$  & $7.6\times 10^{-4}$ \\ 
$30.0$ & $0.013$ & $1.8\times 10^{-5}$ \\ 
\hline
\hline \\[2.mm]
$ $&(b)$ \thinspace \Delta R(\mu,\gamma)>0.7$&$ $\\
\hline
\hline
$E^0_T(\gamma)$ & $W\to\mu\nu\gamma$ & 
$W\to\mu\nu\gamma\gamma$ \\
\hline
$0.1 $ & $3.76$  & $0.071$ \\
$0.3 $ & $2.60$  & $0.037$ \\
$1.0 $ & $1.46$  & $0.012$\\
$3.0 $ & $0.65$  & $2.9\times 10^{-3}$\\
$10.0$ & $0.13$  & $2.0\times 10^{-4}$ \\ 
$30.0$ & $0.012$ & $7.4\times 10^{-6}$ \\ 
\hline
\hline
\end{tabular}
\end{center}
\end{table}

\end{document}